\documentclass[twocolumn]{aastex631}
\usepackage{graphicx}
\usepackage{float}
\usepackage{rotating}
\usepackage{tablefootnote}
\DeclareUnicodeCharacter{2212}{\ensuremath{-}}

\newcommand{\lsim }{{\lower0.8ex\hbox{$\buildrel <\over\sim$}}}
\newcommand{\gsim }{{\lower0.8ex\hbox{$\buildrel >\over\sim$}}}
\newcommand{\Msun}{\ifmmode {M_{\odot}}\else${M_{\odot}}$\fi}
\newcommand{\Lsun}{\ifmmode {L_{\odot}}\else${L_{\odot}}$\fi}
\newcommand{\Rsun}{\ifmmode {R_{\odot}}\else${R_{\odot}}$\fi}

\shorttitle{Star cluster counterparts to the brightest X-ray sources in M51}
\shortauthors{Dage et al.}

\begin{document}

\title{Detecting the Black Hole Candidate Population in M51's Young Massive Star Clusters: Constraints on Accreting Intermediate Mass Black Holes}

\correspondingauthor{Kristen Dage}
\email{kristen.dage@curtin.edu.au}

\author[0000-0002-8532-4025]{Kristen C. Dage}
\affiliation{International Centre for Radio Astronomy Research -- Curtin University, GPO Box U1987, Perth, WA 6845, Australia}

\author[0000-0001-8424-2848]{Evangelia Tremou}
\affiliation{National Radio Astronomy Observatory, Socorro, NM 87801, USA}

\author[0000-0002-1046-1500]{Bolivia Cuevas Otahola}
\affiliation{Departamento de Matem\'aticas-FCE, Benem\'erita Universidad Aut\'onoma de Puebla, Mexico}

\author[0000-0001-8424-2848]{Eric W. Koch}
\affiliation{Center for Astrophysics $|$ Harvard \& Smithsonian, 60 Garden St., Cambridge, MA 02138, US}

\author[0000-0003-1814-8620]{Kwangmin Oh}
\affiliation{Center for Data Intensive and Time Domain Astronomy, Department of Physics and Astronomy, Michigan State University, East Lansing, MI 48824, USA \\}

\author[0000-0002-7092-0326]{Richard M. Plotkin}
\affiliation{Department of Physics, University of Nevada, Reno, NV 89557, USA}
\affiliation{Nevada Center for Astrophysics, University of Nevada, Las Vegas, NV 89154, USA}

\author[0009-0003-7709-5474]{Vivian L. Tang}
\affiliation{Department of Astronomy \& Astrophysics, University of California, 1156 High Street, Santa Cruz, CA 95064, USA}

\author[0009-0004-9310-020X]{Muhammad Ridha Aldhalemi}
\affiliation{Henry Ford College, 5101 Evergreen Rd, Dearborn, MI 48128, USA \\} 

\author[0009-0006-8696-9892]{Zainab Bustani}
\affiliation{Henry Ford College, 5101 Evergreen Rd, Dearborn, MI 48128, USA \\} 
\author[0009-0005-2051-1304]{Mariam Ismail Fawaz}
\affiliation{Henry Ford College, 5101 Evergreen Rd, Dearborn, MI 48128, USA \\}

\author[0009-0006-5976-8120]{Hans J. Harff}
\affiliation{Henry Ford College, 5101 Evergreen Rd, Dearborn, MI 48128, USA \\}

\author[0009-0000-6403-8903]{Amna Khalyleh}
\affiliation{Henry Ford College, 5101 Evergreen Rd, Dearborn, MI 48128, USA \\}

\author[0009-0007-9611-1774]{Timothy McBride}
\affiliation{Henry Ford College, 5101 Evergreen Rd, Dearborn, MI 48128, USA \\}

\author[0009-0006-2324-0738]{Jesse Mason}
\affiliation{Henry Ford College, 5101 Evergreen Rd, Dearborn, MI 48128, USA \\}

\author[0009-0000-8689-3476]{Anthony Preston}
\affiliation{Henry Ford College, 5101 Evergreen Rd, Dearborn, MI 48128, USA \\}

\author[0009-0001-7561-6753]{Cortney Rinehart}
\affiliation{Henry Ford College, 5101 Evergreen Rd, Dearborn, MI 48128, USA \\}

\author[0009-0003-1543-4514]{Ethan Vinson}
\affiliation{Henry Ford College, 5101 Evergreen Rd, Dearborn, MI 48128, USA \\}

\author[0000-0003-1814-8620]{Gemma Anderson}
\affiliation{International Centre for Radio Astronomy Research -- Curtin University, GPO Box U1987, Perth, WA 6845, Australia}

\author[0000-0002-8294-9281]{Edward M. Cackett}
\affiliation{Department of Physics \& Astronomy, Wayne State University, 666 W. Hancock St, Detroit, MI 48201, USA}

\author[0000-0002-9077-6026]{Shih Ching Fu}
\affiliation{International Centre for Radio Astronomy Research -- Curtin University, GPO Box U1987, Perth, WA 6845, Australia}

\author[0000-0003-1814-8620]{Sebastian Kamann}
\affiliation{Astrophysics Research Institute, Liverpool John Moores University, IC2 Liverpool Science Park, 146 Brownlow Hill, Liverpool L3 5RF, UK}

\author[0000-0001-8424-2848]{Teresa Panurach}
\affiliation{Center for Materials Research, Department of Physics, Norfolk State University, Norfolk VA 23504, USA \\}

\author[0000-0003-1814-8620]{Renuka Pechetti}
\affiliation{Astrophysics Research Institute, Liverpool John Moores University, IC2 Liverpool Science Park, 146 Brownlow Hill, Liverpool L3 5RF, UK}

\author[0000-0003-1814-8620]{Payaswini Saikia}
\affiliation{Center for Astro, Particle and Planetary Physics, New York University Abu Dhabi, PO Box 129188, Abu Dhabi, UAE}

\author[0000-0001-9261-1738]{Susmita Sett}
\affiliation{International Centre for Radio Astronomy Research -- Curtin University, GPO Box U1987, Perth, WA 6845, Australia}

\author[0000-0003-1814-8620]{Ryan Urquhart}
\affiliation{Center for Data Intensive and Time Domain Astronomy, Department of Physics and Astronomy, Michigan State University, East Lansing, MI 48824, USA \\}

\author[0000-0002-7383-7106]{Christopher Usher}
\affiliation{The Oskar Klein Centre, Department of Astronomy, Stockholm University, AlbaNova, SE-106 91 Stockholm, Sweden \\}

\begin{abstract}
Intermediate mass black holes ( $10^2 < M_{\textrm{BH}} < 10^5 \Msun$) are an open question in our understanding of black hole evolution and growth. They have long been linked to dense star cluster environments thanks to cluster dynamics, but there are a limited number of secure detections. We leverage existing X-ray observations from \textit{Chandra X-ray Observatory} and optical catalogs from \textit{Hubble Space Telescope} with new radio observations from the \textit{Karl G. Jansky Very Large Array}  to search for any evidence of accreting black holes in young massive clusters in the nearby galaxy M51. We find that of 43 bright ($L_X > 10^{38}$ erg/s) X-ray point sources in M51, 24 had probable matches to objects including possible associated star clusters in the HST Legacy Extragalactic UV Survey catalog, seven of which were classified as contaminants (background galaxies or foreground stars). We explore the optical properties of the remaining 17 sources, including cluster age and mass estimates, and search for radio counterparts in the 8-12 GHz band. The lack of radio counterparts to X-ray sources we know to be associated with young massive clusters in M51 suggests that we do not significantly detect hard-state IMBHs $\sim10^4 \Msun$ or above. However, more sensitive radio facilities like the Square Kilometre Array and next generation Very Large Array may be able to provide evidence for IMBHs with masses down to $\sim 10^3 \Msun$. 

\end{abstract}


\section{Introduction}
\label{sec:intro}
Intermediate mass black holes  (IMBHs; $10^2 < M_{\textrm{BH}} < 10^5 \Msun$) are a crucial stepping stone between stellar mass black holes and the massive black holes observed in centers of galaxies already at early cosmological epochs (\citealt{2018Natur.553..473B}). One of the main theories for supermassive black hole formation is that they are formed from seeds of IMBHs, the observational constraints on IMBHs are few, and their formation channels remain relatively unconstrained \citep{2020ARA&A..58..257G, 2023arXiv231112118A}. Ultraluminous X-ray sources (ULXs; non-nuclear point sources whose X-ray luminosity exceeds the Eddington Limit for a 10 M$_\odot$ BH$-$ around $10^{39}$ erg/s), have been seen as promising IMBH candidates due to their unusually high X-ray luminosity \citep{1989ARA&A..27...87F}, although most are likely an extreme form of a stellar mass compact object accreting at super Eddington rates \citep[][and references therein]{2009MNRAS.397.1836G}.

\begin{table*}
\label{table:allmatches}
\caption{Possible matches between bright X-ray sources in \cite{2023ChJPh..83..579S} with the LEGUS catalog. The ID column refers to ID assigned by \cite{2023ChJPh..83..579S}, the HST ID is the ID assigned by the LEGUS catalog, and we provide the R.A. and Dec along with the F555W and F814W magnitudes, concentration index from LEGUS. We also show the separation in arcseconds and probability of individual match as reported by \textsc{nway}. Src09, src31 and src36 had two different possible counterparts with nearly equal match probabilities and very similar distances.  Src25 and src26, marked with $^+$ are the eclipsing ULXs from \cite{2018MNRAS.475.3561U}. CI is the concentration index reported by LEGUS, and p\_i is the individual match probability.  Column ``Cl1'' indicates the LEGUS classification for the cluster candidate, and column ``Cl2'' is our \texttt{nProFit} classification. }
   \begin{tabular}{lrrrrrrrrrrr}
    ID & HST ID & R.A. \& Dec. & F555W & F814W & CI & Cl1 & Cl2 & Sep (")& p$_i$ (\%) \\ \hline
    src03 & 2141  & 13:29:54.98 +47:09:23& 23.97 $\pm$ 0.07& 24.23 $\pm$ 0.20 & 1.7 &  0 &2& 0.14 & 100 \\
    src04 & 3060  & 13:29:47.48 +47:09:41 & 24.14 $\pm$ 0.13 & 23.67 $\pm$ 0.18 & 2.1 & 0 &0 & 0.39 & 38 \\
    src08 & 7700  & 13:29:46.16 +47:10:42 & 23.37 $\pm$ 0.05 & 22.49 $\pm$ 0.07& 1.6 &  4 &4 & 0.47 & 30 \\
    src09 & 7494  & 13:29:53.31 +47:10:43 & 20.22 $\pm$ 0.05 & 21.57 $\pm$ 0.04 & 1.9 & 4 &4&0.18 &16 \\
    src09 & 7556  & 13:29:53.28 +47:10:43 & 19.16 $\pm$ 0.03 & 18.86$\pm$ 0.03 & 1.5 &  2 &2&0.54 &13 \\
    src10 & 8391  & 13:29:57.61 +47:10:48& 22.99 $\pm$ 0.06 & 22.57 $\pm$ 0.08 & 1.4 & 4 &4& 0.49 &32 \\
    src12 & 8780  & 13:29:49.05 +47:10:55 & 21.97 $\pm$ 0.03 & 21.43 $\pm$ 0.04 & 1.7 & 1&1 & 0.83 & 100 \\
    src16 & 11819 & 13:29:53.56 +47:11:27 & 21.78 $\pm$ 0.03 & 21.15 $\pm$ 0.05 & 1.5& 4 &2& 0.09 & 35 \\
    src17 & 12622 & 13:29:53.58 +47:11:33 & 22.91 $\pm$ 0.27 & 23.20 $\pm$ 0.45 & 2.0 & 0 &0& 0.55 & 59 \\
    src18 & 13341 & 13:29:43.31 +47:11:35 & 22.98 $\pm$ 0.04 & 22.83 $\pm$ 0.08 & 1.4 & 4 &2& 0.07 & 30 \\
    src19 & 13586 & 13:29:54.77 +47:11:36 & 22.75 $\pm$ 0.04 & 22.32 $\pm$ 0.18 & 1.8 & 4 &2& 0.34 & 30 \\
    src20 & 13003 & 13:29:54.23 +47:11:37 & 23.24 $\pm$ 0.20 & 22.76 $\pm$ 0.27 & 2.1 & 4 &4& 0.59 & 43 \\
    src21 & 14843 & 13:29:55.86 +47:11:45 & 22.41 $\pm$ 0.04 & 21.91 $\pm$ 0.05 & 2.0 & 4 & 4&0.25 & 28 \\
    src22 & 16025 & 13:29:45.6s +47:11:51 & 24.26 $\pm$ 0.06 & 24.33 $\pm$ 0.19 & 1.5 & 0 & 0&0.90 & 100 \\
    src23 & 16595 & 13:29:50.66 +47:11:55 & 23.65 $\pm$ 0.10 & 23.27 $\pm$ 0.12 & 1.6 & 0 & 0&0.40 & 22 \\
    src24 & 18061 & 13:29:57.65 +47:12:07 & 22.63 $\pm$ 0.03 & 21.99 $\pm$ 0.04 & 1.9 & 2 & 2&0.27 & 100 \\
    src25$^+$ & 21034 & 13:29:39.96 +47:12:36 & 23.65 $\pm$ 0.04 & 23.85 $\pm$ 0.09 & 1.5 & 0 &3 & 0.17 & 100 \\
    src26$^+$ & 21770 & 13:29:39.46 +47:12:43 & 22.60 $\pm$ 0.03 & 22.53 $\pm$ 0.04 & 1.4 & 0 &4& 0.20 & 66 \\
    src28 & 22030 & 13:30:07.85 +47:12:46 & 23.86 $\pm$ 0.06 & 23.89 $\pm$ 0.12 & 1.4 & 0 & 2&0.62 & 100 \\
    src31 & 25510 & 13:30:04.32 +47:13:21 & 23.49 $\pm$ 0.06 & 22.85 $\pm$ 0.06 & 1.6 & 0 & 2&0.25 & 25\\
    src31 & 30129 & 13:30:04.28 +47:13:21& 23.12 $\pm$ 0.06 & 22.71 $\pm$ 0.06 & 1.4 & 4 & 4 &0.36 & 24\\
    src34 & 26467 & 13:29:58.36+47:13:32& 23.92 $\pm$ 0.06 & 22.61 $\pm$ 0.06 & 1.7 & 0&2 & 0.61 & 100 \\
    src35 & 27115 & 13:29:38.64+47:13:36 & 15.50 $\pm$ 0.02 & 14.99 $\pm$ 0.03 & 2.1 & 0&4 & 0.18& 100 \\
    src36  & 27626 &  13:30:00.99+47:13:44& 23.38 $\pm$ 0.13 & 22.57 $\pm$ 0.12 & 1.5 & 4&4 & 0.20 & 12 \\
    src36 & 27620 & 13:30:00.98+47:13:44 & 23.22 $\pm$ 0.08 & 22.39 $\pm$ 0.06 & 1.8 & 3 &4& 0.31 & 12 \\
    src40 & 27137 & 13:29:56.20+47:14:51 & 23.91 $\pm$ 0.04 & 23.42 $\pm$ 0.06 & 1.5 & 0 &2& 0.45 & 100 \\
    src43 & 20431 & 13:29:57.46+47:16:11 & 22.14 $\pm$ 0.03 & 21.08 $\pm$ 0.04 & 1.5 & 1 &1& 0.41 & 100 \\ \hline
   \end{tabular}

\end{table*}

\begin{figure*}
    \centering
    \includegraphics[width=7in]{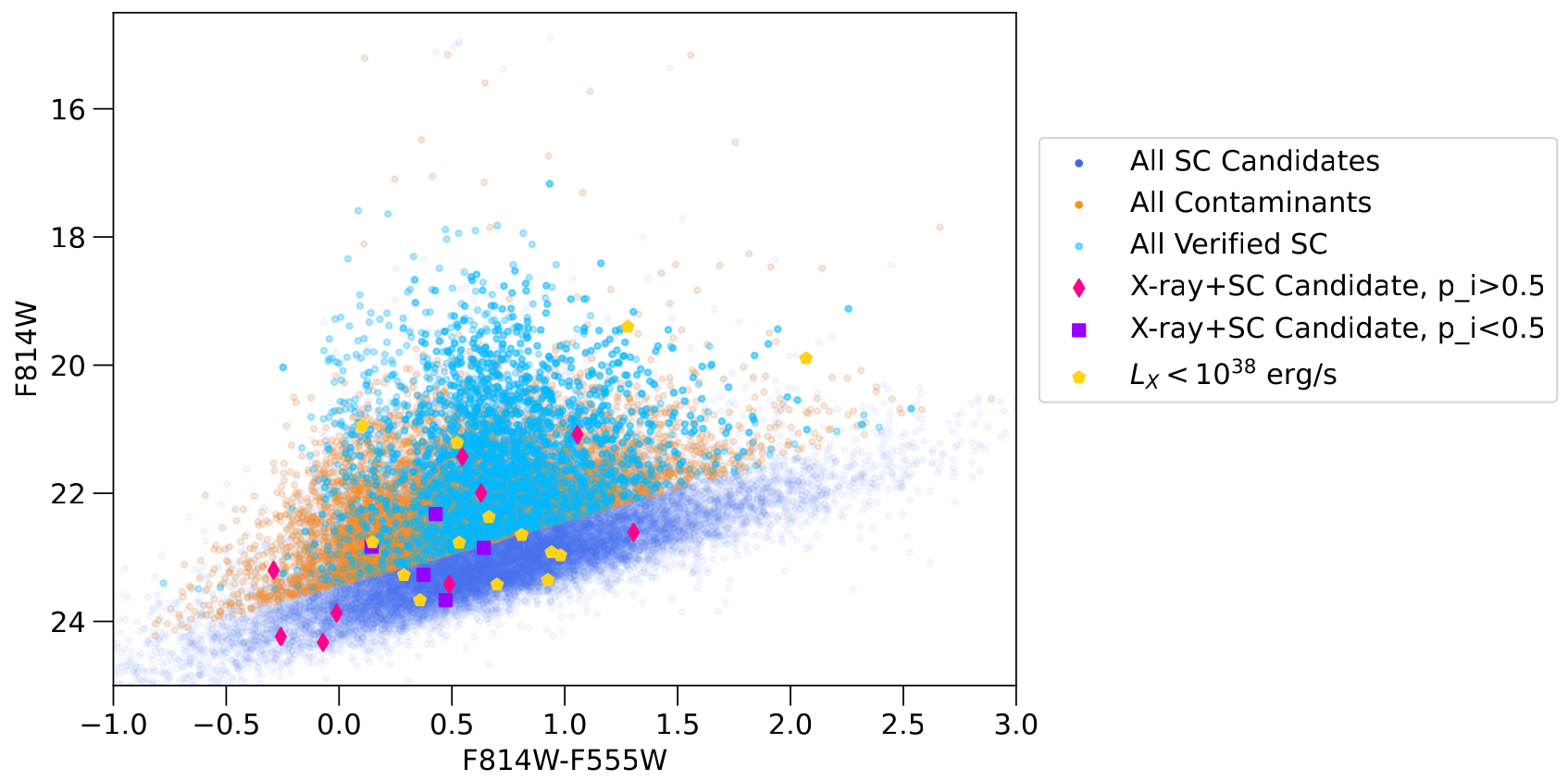}
    \caption{Color-magnitude diagram of all LEGUS cluster candidates, along with the verified star clusters (SCs) and contaminants. The yellow pentagons represent the X-ray sources that matched to  high probability ($>$ 50 \%) matches to cluster candidates, lower probability ($<$ 50 \%) matches to cluster candidates, or contaminants.}
    \label{fig:cmd}
\end{figure*}

Our understanding of ULXs evolved significantly when X-ray pulsations were detected from M82-X2 \citep{2014Natur.514..202B}, the first of several observations to show that neutron stars could accrete well above their own Eddington Limit. However, these pulsations are transient, which makes it difficult to rule out a neutron star primary from timing observations \citep{2014Natur.514..202B,2019ApJ...875..144P}.
We thus know that an observed X-ray luminosity of $10^{39}$ erg/s can be produced by a massive ($>10 \Msun$) black hole accreting at a sub-Eddington rate, or super-Eddington accretion onto a stellar mass object, either a neutron star or a $<$10 M$_\odot$ BH. 

 From the X-ray perspective,  ULXs have almost exclusively been found outside of the Milky Way, with most located in spiral galaxies \citep[e.g.,][and many references therein]{2020MNRAS.498.4790K}. A small number of ULXs have been identified in globular clusters associated with elliptical galaxies \citep[][and references therein]{2020MNRAS.497..596D}. While globular clusters were initially looked to as the natural birthplaces of IMBHs, radio studies like that of \cite{2018ApJ...862...16T} have placed relatively stringent constraints on the presence of massive IMBHs in Galactic globular clusters, although the presence of IMBHs in systems like $\omega$ Cen and 47 Tuc are currently being debated \citep[and references therein]{2024ApJ...961...54P,2024Natur.631..285H}, and evidence for IMBHs have been found in tidally stripped nuclei \citep[][and references therein]{2022ApJ...924...48P}. By comparison, young massive clusters remain fairly poorly studied.

Much theoretical work has also focused on the formation of BHs in young, massive star clusters$-$particularly the formation of IMBHs. Young massive star clusters (YMCs) are typically younger than globular clusters, with ages from 1 to 100s of Myr and masses greater than $10^4$ M$_{\odot}$. They can however overlap in size with globular clusters \citep[e.g.,][]{2010ARA&A..48..431P}.  As dense stellar environments, YMCs are conducive towards forming massive BHs via runaway stellar collisions and mergers, as shown by \textsc{N-BODY} simulations \citep[e.g.,][]{2002ApJ...576..899P,2004Natur.428..724P,2008MNRAS.383..230M, 2014ApJ...794....7M, 2019MNRAS.487.2947D,2021MNRAS.507.3612R,2021MNRAS.507.5132D}.

However, unlike globular clusters, the observations of X-ray sources in YMCs are relatively sparser: \cite{2011ApJ...741...86R,2012ApJ...758...99R} found observational connections between star clusters and X-ray binaries, in NGC 4449 and the Antennae galaxies, and some ULXs have been serendipitously identified in YMCs \citep[such as][]{2006ApJ...645..264T, 2007ApJ...668..124A,2014MNRAS.442.1054H,2016ApJ...828..105A,2017MNRAS.469..671L,2018MNRAS.475.3561U}. Subsequent studies by \cite{2023MNRAS.522.5669B,2024MNRAS.529.1507A} targeted M31 and M33, NGC 4490 and NGC 4214. \cite{2023MNRAS.522.5669B} found that while  X-ray binaries (XRB) do show spatial correlation with young star clusters, very few of the brightest XRBs are observed within the young star clusters. \cite{2024MNRAS.529.1507A} found that a high percentage of detected XRBs are associated with star clusters, with most being in the younger and less massive clusters. \cite{2023ApJ...953..126H} explored the connection between XRBs and star clusters in six star forming galaxies, and found that the youngest XRB hosting clusters in their sample were the more massive clusters.

\begin{table*}

\caption{Age and mass estimates of cluster candidates, along with peak X-ray luminosity as reported by \cite{2023ChJPh..83..579S}, along with their characteristic best-fit X-ray spectral shape (determined by the model with the best-fit statistics over most observations, `DB' for a disk blackbody best fit, `PL' for a power-law model  best fit). The ``Variable'' column has t for sources with only one X-ray detection in ObsID 1622, a 30 ks observation taken in 2001, and are not observed above the detection threshold in subsequent observations of equal or greater sensitivity,  N for sources which do not vary significantly within their uncertainties, M for moderately variable sources (which vary within their uncertainties but not by an order of magnitude), and H for highly variable sources which vary by an order of magnitude. Src 25 and src26 are the eclipsing ULXs from \cite{2018MNRAS.475.3561U}. Four sources did not have a mass or age estimate, so we assume the lowest age and mass estimates from the non-contaminated LEGUS list as upper limits age and masses for these sources. }
\label{table:xray2}
\begin{tabular}{lrrrrrr}
ID & Age (yr) & Mass (\Msun)  & p$_i$ (\%) & Peak $L_X$ (erg/s) &Variable&X-ray spectrum\\ \hline
src03 $^*$& $<$1e6&  $<$84 & 100 & 1.44 $\times 10^{38}$ &M &DB\\
src04 & 1e8 & 1542 & 39 & 7.24  $\times 10^{38}$ &M&PL\\
src12 & 2e8 & 17800 & 100 & 5.12  $\times 10^{38}$& t&PL \\
src16 & 4e6 & 10670 &35 &1.22$\times 10^{39}$&N&PL\\
src17 &  $<$1e6 &  $<$84 & 59 &  6.02  $\times 10^{38}$&N&DB\\
src18 & 1e6&650.7 & 30 & 6.91   $\times 10^{40}$& M&DB\\
src19 & 7e8 & 19450 & 30& 9.77 $\times 10^{38}$& N&PL\\
src22 & 1e6 & 292 & 100 & 2.75  $\times 10^{38}$&N &DB\\
src23 & 2e6 & 584 & 23 & 1.20  $\times 10^{40}$&M&DB \\
src24 & 2e8 & 13930 & 100 & 1.02  $\times 10^{39}$ &N&DB? \\
src25 $^+$ &  $<$1e6 &  $<$84 & 100 &  5.88 $\times 10^{39}$&M&PL\\
src26 $^+$ & $<$1e6&  $<$84 & 66 & 1.82  $\times 10^{39}$  &H&PL?\\
src28 & 3e6 & 230 & 100 & 2.19  $\times 10^{38}$ &N &
PL\\
src31 & 2e6 & 1101 & 25 & 1.41  $\times 10^{39}$ &H&PL\\
src34 & 4e7 & 6107 & 100 & 3.47  $\times 10^{38}$ &M&DB?\\
src40 & 6e6 & 316 & 100 &  6.45 $\times 10^{38}$ & t&PL\\
src43 & 2e9 & 84860& 100 & 1.20 $\times 10^{39}$ & M &DB? \\ \hline
\end{tabular}
\end{table*}

\begin{figure*}
    \centering  \includegraphics[width=0.9\linewidth]{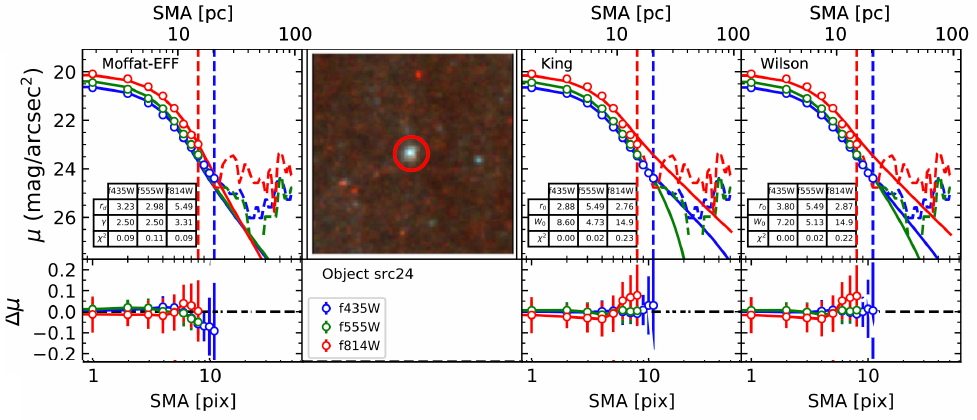}
    \caption{ Surface brightness profiles (SBPs) of src24 in the F435W, F555W and F814W HST observations in blue, green and red empty circles, obtained by {\sc nProFit} from isophotal fitting performed to the image centered at src24 coordinates (shown in the lower panels from left to right).  The SBPs are fitted using Moffat-EFF (left-most panel), King (third panel from left to right), and Wilson (right-most panel), fitted to the SBPs up to the fitting radius shown in vertical lines, following the same color code, with the minimum fitting radius between all bands shown as a circle in the snapshot in the second upper panel from left to right. The basic structural parameters obtained for each model are summarized in the inset tables. The bottom panels show the residuals of the dynamical models fits performed by {\sc nProFit}.}
    \label{fig:sbp_prof_fit}
\end{figure*}
The XRBs in all of these studies span a wide range in X-ray luminosity from $10^{36}$ erg/s to $10^{39}$ erg/s, and may be an assortment of high mass X-ray binaries (HMXBs), intermediate mass X-ray binaries (IMXBs) and low mass X-ray binaries (LMXBs). From an observational perspective, it is not possible to classify the nature of the compact object from the X-ray luminosity alone; as previously discussed, neutron stars are able to achieve X-ray luminosities above $10^{39}$ erg/s, and the same is true of stellar mass black holes \citep{2009MNRAS.397.1836G}. However, this is an area where radio follow up may be able to differentiate between a stellar mass accretor, and a more massive one. Radio observations of ULXs are key to determining whether the accretor is an intermediate mass black hole, or a stellar mass neutron star or black hole \citep{2013MNRAS.436.3128M, 2015MNRAS.448.1893M,panurach2024neutronstarultraluminousxray}.  If a $10^4$ M$_\odot$ IMBH ($L_X\approx10^{39}-10^{40}$ erg s$^{-1}$),  is accreting at a very low Eddington ratio ($<0.01$ $L_{\rm Edd}$), it could launch a compact radio jet, as observed from other low-luminosity active galactic nuclei \citep{2008ARA&A..46..475H}. 
sing the fundamental plane of black hole activity set forth in \cite{Merloni03, Falcke04}. The fundamental plane of black hole activity is a non-linear empirical relationship observed in both hard-state stellar-mass black holes and their supermassive counterparts, linking the X-ray (or nuclear [OIII] emission line) luminosity, compact radio luminosity, and the mass of the black hole, demonstrating a consistent connection across vastly different black hole scales \citep{Merloni03, Falcke04,2015MNRAS.450.2317S,2018A&A...616A.152S}. 

\begin{figure*}
\includegraphics[width=6in]{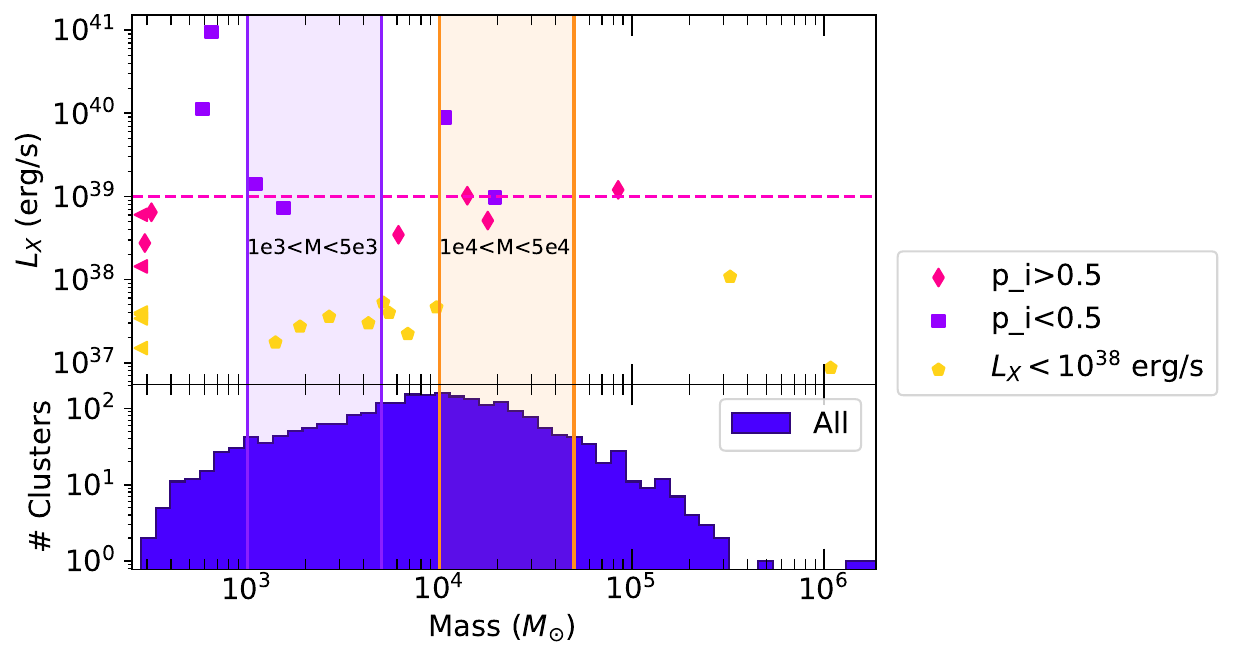}

\caption{X-ray luminosity versus masses of cluster candidates. The upper panel show the estimated mass of the cluster candidate versus the X-ray luminosity. The lower panel shows a histogram of mass estimates for all of the cluster candidates (with contaminants removed), and we have shaded in regions of clusters with masses between 1000 and 5000, and 10000 and 50000 to better compare to \cite{2021MNRAS.507.5132D}. } 
\label{Fig:mass_age2}
\end{figure*}
At the distances of nearby galaxies like M51 (D$<$ 10 Mpc), the jetted radio emission from IMBHs can be readily observed by the \textit{Karl G. Jansky Very Large Array} (VLA) \citep{2011ApJ...739L...1P}.
We take advantage of the wealth of archival \textit{Chandra} \citep{Weisskopf02} X-ray observations of the high star formation rate spiral galaxy M51, known to host a multitude of ULXs and other X-ray bright sources, along with the publicly available star cluster catalogs from the HST Legacy Extragalactic UV Survey (LEGUS; \citealt{2015AJ....149...51C,2017ApJ...841..131A}), along with new radio observations from the VLA to assess whether any of the XRBs associated with young massive clusters could possibly be IMBHs more massive than $10^4 \Msun$. 
We will be focusing on the brightest X-ray sources in M51 ($L_X > 10^{38}$ erg/s if at the distance of M51 (8.58 Mpc, \citealt{2016ApJ...826...21M}). We perform Bayesian cross matching analysis between the X-ray and optical using \textsc{nway} \citep{2018MNRAS.473.4937S}, and compare our observations to state-of-the-art simulations of young star clusters.  We also discuss what range of IMBH masses are actually observable in X-ray and radio at the distance of M51, given current observational facilities.


\section{Data and Analysis}
We combine multiwavelength observations from \textit{Chandra}, \textit{HST} by performing a Bayesian cross match between \textit{Chandra} X-ray point sources and optical catalogs from LEGUS\footnote{\url{https://legus.stsci.edu/}}, and search for a radio counterpart from the VLA to any X-ray bright ($L_X > 10^{38}$ erg/s) sources with a candidate cluster optical counterpart. There are a number of caveats associated with our process and we list them below. However, despite the many challenges, our intention is to determine the maximum number of potential accreting IMBHs in star clusters to provide useful benchmarks to compare theory to, and thus we present all probabilities and caveats to probe both the optimistic and the realistic scenarios.

\subsection{X-ray analysis}

M51 has been studied in X-ray by \textit{Chandra} 23 times in the last 21 years, covering several different fields of the galaxy, spanning about 4.5 arcmin. We did initial cross matching with the Chandra Source Catalog V. 2.0 (CSC, \citealt{Evans10}) to have a reasonably complete catalog of the entire field. We note that the CSC did have some (about 1\%) spurious detections at the lower luminosity ends, and the reported flux values differ from published detailed spectroscopic studies, perhaps due to the innate variability of some of the sources, or assumptions about spectral shape \footnote{\url{https://cxc.cfa.harvard.edu/csc/caveats.html}}. For the brightest sources, we obtain X-ray luminosities and spectral parameters from existing long-term studies of M51's X-ray binaries which perform detailed X-ray spectroscopy \citep{2023ChJPh..83..579S}. To target the faintest detectable X-ray binaries, we run source detection and flux estimation on the longest available observation.

To supplement the findings of the CSC and \cite{2023ChJPh..83..579S} on M51’s brightest X-ray sources, we processed the longest available  observations of M51, ObsIDs 13812 (2012-09-12, 160ks), 13813 (2012-09-09, 180ks), 18314 (2012-09-20, 190ks), 13815 (2012-09-23, 68ks) and 13816 (2012-09-26, 74 ks) observed in 2012 (PI: Kuntz). The observations from 2012 are much longer than any other available observation, and are useful to find X-ray sources at the faintest detection limit for the sake of completeness. 
For this analysis, we used the {\it Chandra} Interactive Analysis of Observations \citep[CIAO, version 4.15.1,][]{Fruscione06}, with the latest calibration files from the \textit{Chandra} Calibration Database (CALDB). We reprocessed all the data using \texttt{chandra\_repro}, and merged them with \texttt{merge\_obs} tool.

 While we cannot perform detailed spectroscopy due to the low number of source counts ($<10$), nor trace the long-term X-ray behavior of the faint sources, we used \texttt{wavdetect} to identify fainter X-ray sources down to L$_X \sim 10^{37}$ erg/s. We used a range of spatial scales of 2, 4, 8, 16, 24, 32, 48 pixels with a significance threshold of $10^{-6}$. This corresponds to about 1 false alarm per $1024 \times 1024$ pixel image. 
We computed unabsorbed X-ray fluxes  using the \texttt{srcflux} tool, probing the 0.3--10.0 keV energy range. We assumed a hydrogen column density of $3.8 \times 10^{18}\, \text{cm}^{-2}$, which we obtained from \cite{2023MNRAS.521.2719Y}.  We assumed a fixed power-law model with a photon index of 1.7, and focused on the full 0.3--10.0 keV energy band to be consistent with \cite{2023ChJPh..83..579S}.  Details of the X-ray and star cluster properties are in Table \ref{table:xray2}.

\subsection{Star Cluster Catalog and Bayesian Cross Matching Analysis}

According to the CSC, there are less than 200 X-ray point sources in a 5 arcminute radius around M51. In a similarly sized region, LEGUS reports over 30,000 optical point sources. As detailed in \cite{2023ApJ...953..126H}, the probability of chance superposition is certainly not negligible. By using \textsc{nway}\footnote{\url{https://github.com/JohannesBuchner/nway}}, \citep{2018MNRAS.473.4937S}, software designed for Bayesian cross-matching of multiple catalogs, we can provide some measure of constraint on the probability of the individual cluster candidate being a match to a given X-ray source, as \textsc{nway} does compute the probability for a random chance alignment of two unrelated sources and folds this into the probability of association.  \textsc{nway} reports two probabilities, p\_any, the probability that there is a counterpart in the matching catalog and p\_i, the probability that a given source is an individual match. We only selected sources with a high p\_any ($> 0.5$ \citep{2018MNRAS.473.4937S}. We report the highest p\_i values; sources with a very high p\_i had one close match with in the uncertainty radius, and sources with lower p\_i had multiple matches. We report all matches with p\_i$>0.1$, which show up as duplicate matches, e.g. in the case of src98, src31 and src36.

Another word of caution about the cross-matching is that the LEGUS catalog of 30,000 objects is neither the full photometric list, nor the full star cluster candidate catalog. It is a photometric list of any source with a concentration index (see Section 2.35) greater than 1.35, with likely contaminants denoted. However, for our purposes, we are content to view the matches as the maximum number of X-ray sources that could \textit{possibly} be matched to a cluster candidate.

We performed Bayesian cross matching using \textsc{nway} to cross match the full LEGUS cluster candidate catalog and the CSC, with a maximum cross match radius of 1.1 arcseconds. The full LEGUS catalog contains 30,176 objects. Of these, 10,925 met the cutoff to be considered for classification, with 7551 as class 4 contaminants, 385 class 3 (compact stellar sources), and 2989 class 1 or 2 clusters (most likely to be clusters). The remaining 19251 objects are unclassified class 0 objects. These classifications are discussed in further detail in Section 2.32.

\begin{figure*}

\includegraphics[width=6in]{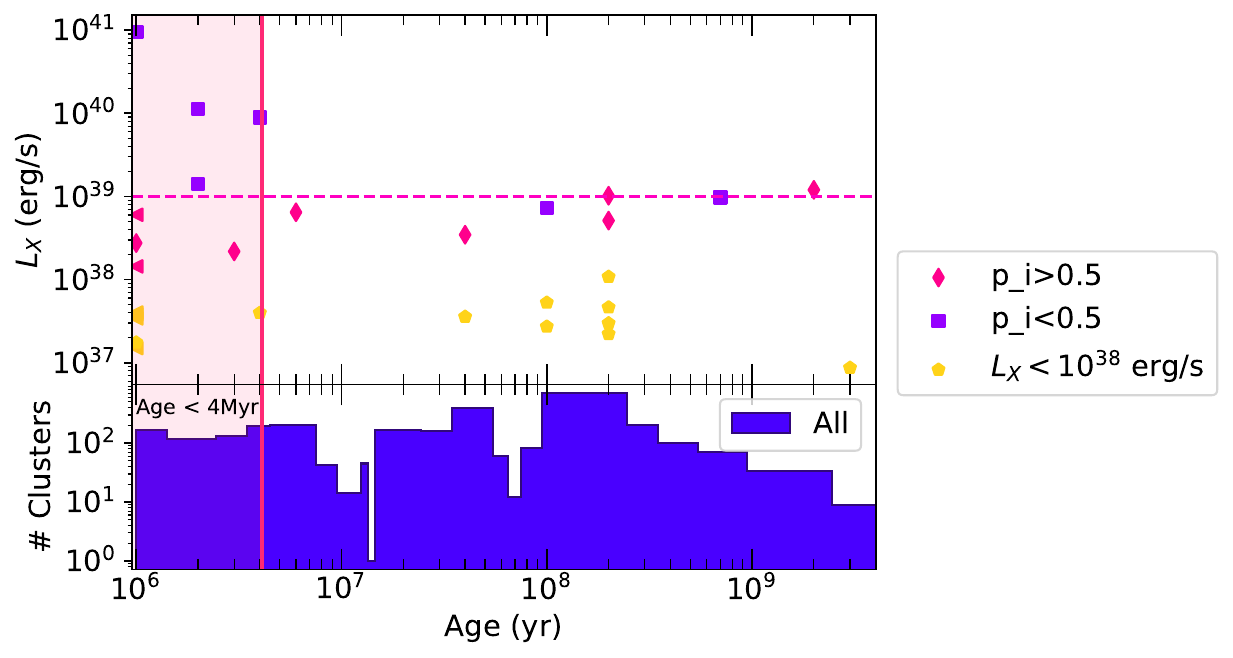}

\caption{X-ray luminosity versus ages of cluster candidates. The upper panel show the estimated age of the cluster candidate versus the 0.3-10 keV X-ray luminosity. The lower panel shows a histogram of age estimates for all of the cluster candidates (with contaminants removed). For visualization purposes, we shaded in ages of less than 4 Myr, where it is possible that enough ICM may be present for a black hole to accrete without a companion.} 
\label{Fig:mass_age}
\end{figure*}
Thanks to \cite{2023ChJPh..83..579S}, we were able to identify the brightest X-ray sources which are accompanied by detailed spectral modeling (both with a single power-law index and a single blackbody disc). Their analysis also includes the long-term properties of the source, to look for both variations in X-ray luminosity and X-ray spectral shape. We found that of the 43 bright X-ray sources in \cite{2023ChJPh..83..579S}, 24 of them match to a source in the LEGUS catalog. Of these, 8 had high probability matches with a class 4 object, and one was clearly a foreground star (src35). 

\textsc{nway}, in addition to reporting the match probability, also provides match rankings, where in the case of low probability matches, other probable matches are also provided. For all of the cluster candidate (class 0,1,2 or 3) sources with match probabilities less than 50\%, we examined the probability values and separations. All of the second rank matches had much lower probability and much farther in separation than the first ranked match, and we do not report them. The match probabilities, separations, source ID from \cite{2023ChJPh..83..579S}, and properties measured by LEGUS are reported in Table \ref{table:allmatches}.  Figure \ref{fig:cmd} shows the color-magnitude diagram of the full LEGUS catalog, the verified star clusters, and verified contaminants, along with the potential cluster matches to the bright X-ray sources. We also compared our stacked X-ray catalog, and Table \ref{table:appendix} in the Appendix shows the low X-ray luminosity matches and cluster properties.

 According to the CSC, the absolute astrometric positional uncertainty at the 90\% level is 1.1 arseconds \footnote{\url{https://cxc.harvard.edu/cal/ASPECT/celmon/}}. As outlined in \cite{2015AJ....149...51C}, the WCS for the LEGUS images have been aligned to a WFC3/UVIS image because the coordinates were derived from the more accurate Guide Star II catalog. 
 While we are content to proceed with searching for matches within the 1.1 arcsecond positional uncertainty of \textit{Chandra}, we want to emphasize that this search is on a population basis, and recommend performing more detailed astrometry (e.g. \citealt{2024MNRAS.52710185A}) for any individual source of interest. 

\subsection{Young Star Cluster Classification \& Star Cluster Structural Parameter Modeling}

As detailed in \cite{2017ApJ...841..131A}, the young star cluster catalogs are obtained from HST observations in the following manner: the concentration index (CI) is calculated by obtaining the magnitude difference of each source at 1 and 3 pixel aperture radii to select out objects with more concentration of light from the rest. Their pipeline performs multiband photometry for these sources along with averaged and CI-based aperture correction. To estimate ages and masses of the clusters, the spectral energy distribution of each source is fitted with single stellar population models, both from the Padova and Geneva stellar libraries \citep{1999ApJS..123....3L,2005ApJ...621..695V}. Three different types of internal extinction are applied, a Milky Way extinction law, and two different types of starburst extinction models. We used the default reference catalog which uses aperture based correction, the Padova stellar library, and the Milky Way extinction. 
Cluster classification is based on a combination of both visual inspection by members of the team, and machine learning, with classification quality flags being included in the full catalog. Classification is only performed on sources detected in at least four bands and brighter than -6 V mag and photometric errors of 0.3 mag or less. Missed clusters are added back in at the visual inspection stage, typically below 1\%. The flags are 0,1,2,3, and 4, with 0 being unclassified sources and 4 being contaminants (including supernova remnants, foreground stars, background galaxies, and spurious detections). Class 1 and 2 sources are the most likely to be clusters, and class 3 are either less compact clusters, or compact associations. 

Only three of our bright X-ray sources matched to a class 1 or 2 source, and we found a low-probability match to a class 3 source (Table \ref{table:allmatches}). The majority of our sources matched unclassified (class 0) star cluster candidates. For the rest of this analysis, we heavily emphasize that these are cluster candidates, and that the masses and ages provided should be treated as estimates.

\begin{figure*}
    \centering
    \includegraphics[width=7in]{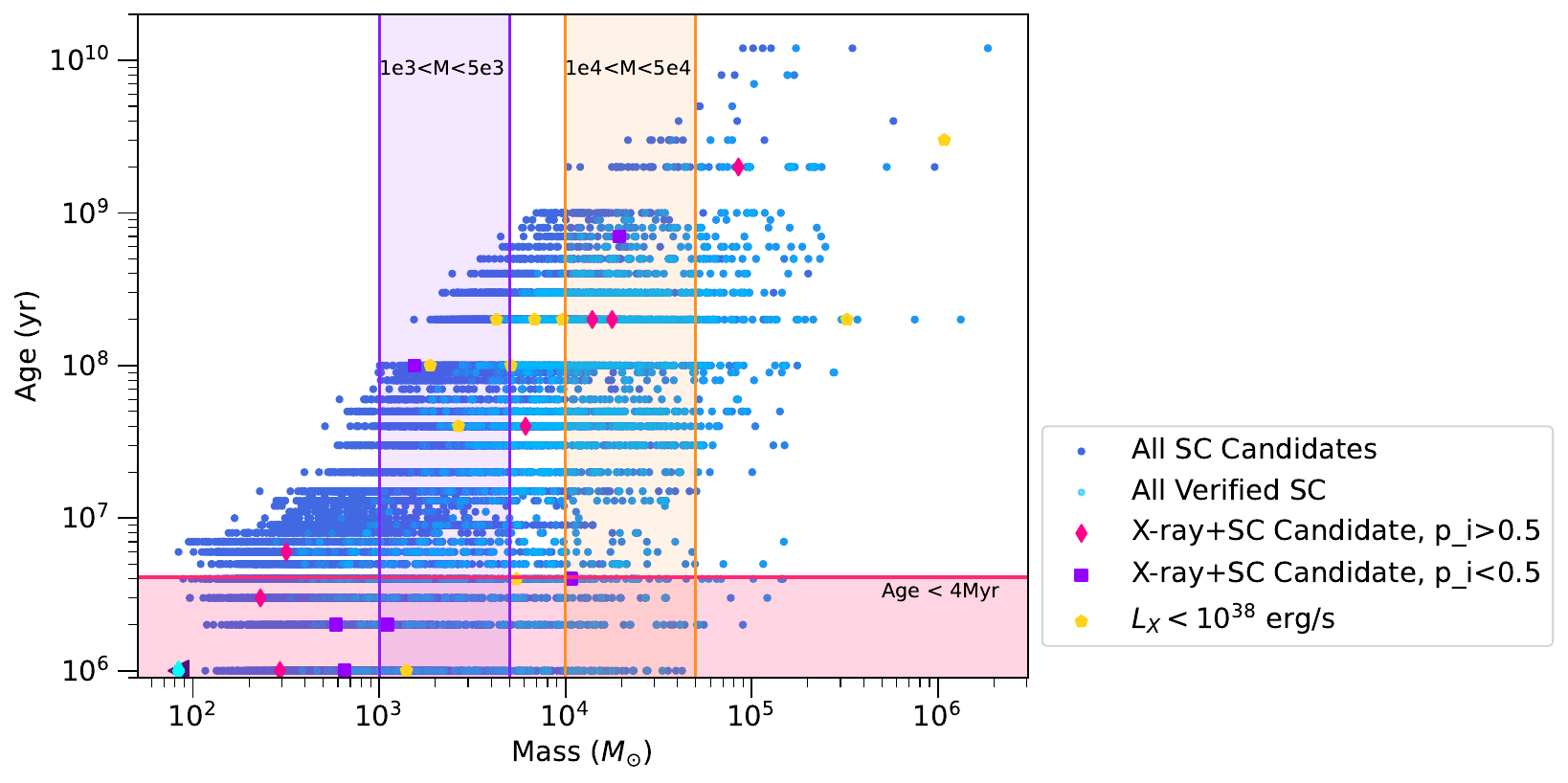}
    \caption{Mass versus age estimates of all LEGUS star cluster candidates (with class 4 objects removed). Verified star clusters are overlaid in a lighter color. The solid pink diamonds represent the properties of the star cluster candidates with high ($>$ 50\%) probability matches, and the purple squares represent the star cluster candidates with low ($<$ 50\%) probability matches. The yellow pentagons are the $L_X <10^{38}$ erg/s source, which are all higher probability matches to candidate star clusters (p\_i $> 50\%$). We highlight the same regions of interest as before, with age less than 4 Myr, and cluster masses between 1000 and 5000 $\Msun$, and cluster masses between 10000 and 50000 $\Msun$.  }
    \label{fig:age_mass}
\end{figure*}
To ensure more secure cluster classification, we used the n-Profile Fitting Tool  \texttt{nProFit} \citep{2022cuevas} to fit the observed Surface Brightness Profiles (SBPs) of the cluster candidates described previously. \texttt{nProFit} is a publicly available code that enables fitting dynamical PSF-convolved models \citep[King, Wilson, and Moffat-EFF,][]{1966King,1975wilson,1987elson} to the observed SBPs of star clusters. Such a procedure is carried out by \texttt{nProFit}, estimating and subtracting in the first place a local background value from images trimmed and centered at each cluster candidate position. Subsequently, the SBPs are extracted by \texttt{nProFit} from the background-subtracted images by means of isophotal fittings, which are performed considering the ellipticity values computed by \texttt{nProFit} in a previous iteration of the isophotal fitting. We show an example of the performed fits in Figure \ref{fig:sbp_prof_fit}, where we illustrate the fitting of Moffat-EFF, King and Wilson models to src24, with Moffat-EFF the best-fit model in the F814W band. 
Considering the obtained structural parameters, namely radii ($r_0$ or $r_d$) and shape ($W_0$ or $\gamma$) parameters, we assigned a classification to each cluster candidate, using the same flags used in LEGUS catalog (where for us 0 is too faint to fit properly, 1,2, and 3 are relatively compact and 4 is extended), and compared both classifications. We found that our classification matches LEGUS classification in 50\% of the cases. We highlight that we correctly identified the sources identified as clusters in LEGUS (src12, src24 and src43). In addition, we identified 8 sources as compact and diffuse clusters , that were identified in LEGUS as spurious detections (src3, src16, src18, src19, src28, src31, src34, and src40). The rest of the clusters were identified as spurious sources or sources below the selection cut defined in LEGUS. For the rest of this analysis, we proceed with sources we classify as 1 or 2, and we also consider the ambiguous (class=0) cases.

\subsection{Radio Data}
Four quadrants of M51 were observed by the VLA observations between June-July 2023 (NRAO/VLA Program ID 23A-104, PI: K.Dage) in the most extended A-array configuration. The data were taken with X-band receivers (8-12 GHz). We used the 3-bit samplers, with two independent 2048-MHz wide basebands centered at 9.0 GHz and 11.0 GHz. The bandwidth was divided into 128-MHz wide spectral windows, and each spectral window was sampled by 64 channels.  All observations were obtained in full polarization mode. Observations typically alternated between 6.5 min on target and 1 min on a phase calibrator (J1335+4542). 3C286 (1331+305) was observed at the start of each block as bandpass and flux calibrator. The total time of each observing block was 2 hours long with typically 1.1 hours on source. Four different fields of M51 were targeted centered at: \\ 
RA: 13h29m43.309s , Dec: 47°11'34.930" \\ 
RA: 13h30m7.547s, Dec: 47°11'6.270" \\
RA: 13h30m6.001s, Dec: 47°15'42.480"  \\
RA: 13h29m50.656s, Dec: 47°11'55.206", 
respectively. 

The data were calibrated and imaged following the standard procedures with the Common Astronomy Software Application \cite[CASA;][]{2007casa,2022casa}.  
Briggs weighting scheme \cite[robust=0,][]{1995briggs} and frequency dependent clean components (with two Taylor terms; nterms=2) were used in imaging to mitigate large-bandwidth effects \citep{2012rau}.  
The mean rms noise of each primary beam corrected image was 4~$\mu$Jy/beam and the median synthesized beam in the images is 0.16" $\times$ 0.14". This corresponds to a cut-off 3 $\sigma$ radio luminosity of 1.06 $\times 10^{34}$ erg/s at 10 GHz.

No significant radio counterpart was detected for any of the X-ray sources, except for the soft X-ray source Src03, which had a nearby radio counterpart ($<$ 1 arcsec) that was 70 $\pm 8.4 \mu$Jy. For the sources not detected in radio, and in the X-ray hard state, we can suggestively use these upper limits (beholden to the many assumptions underlying the fundamental plane) to exclude evidence for IMBHs of masses greater than $10^4 \Msun$. As outlined in \cite{panurach2024neutronstarultraluminousxray}, radio continuum detection is critical towards disentangling whether the X-ray emission could be produced by an IMBH, as evidence for radio jets would be observed.

\section{Results and Discussion} \label{sec:results}
We leveraged archival \textit{Chandra} X-ray and \textit{HST} optical observations of M51, along with new radio observations from the \textit{VLA}. We found that of the 43 brightest X-ray sources in M51, 23 matched to cluster candidates in the LEGUS catalog, 8 of which were contaminants. Of the remaining 15, 3 had high probability matches to a classified star cluster, 8 had high probability matches to an unclassified star cluster candidate and 3 had low probability matches to a star cluster candidate. We also point out Src36, a known neutron star ULX \citep{2020ApJ...895...60R}, which had equally low probability matches to a contaminant or an unclassified star cluster. After using \texttt{nProFit}, we conclude that neither potential optical counterpart is a star cluster. Overall, we found 24 bright X-ray sources possibly matching to an optical counterpart in the LEGUS catalog, and 20 low luminosity sources ($L_X < 10^{38}$ erg/s) with high probability ($>$ 50\%) individual matches.

\begin{figure}
    \centering
    \includegraphics[width=3.5in]{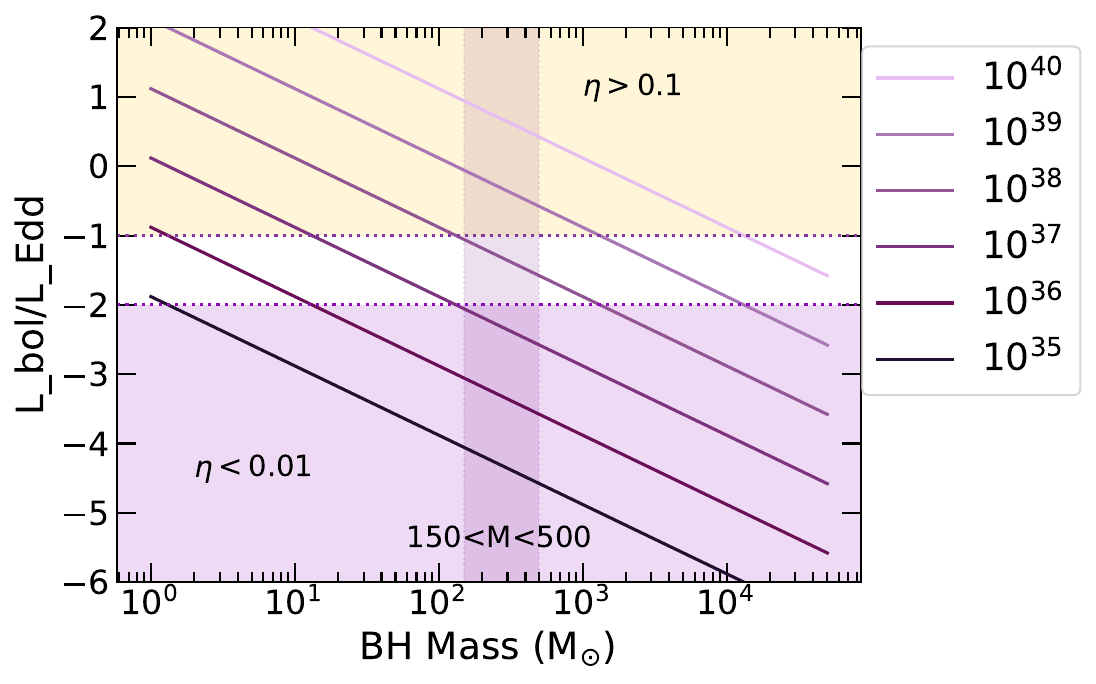}
    \caption{Accretion efficiency for black holes from 1 $\Msun$ to $5\times 10^{4}\Msun$, with a range of X-ray luminosities from $10^{35}-10^{40}$ erg/s. If $\eta$ is below 0.01, then the assumptions of the fundamental plane hold, and one can expect radio emission from the source. If $\eta$ is above 0.1, then the source is expected to be in the super Eddington state.}
    \label{fig:accretioneff}
\end{figure}
An infrared study by \cite{2017MNRAS.469..671L} suggests that Src09 may have a star cluster counterpart. In Table \ref{table:allmatches} we list the highest probability match for Src09, which is a class 4 object, as well as the match probability and separation for the only classified cluster within the 1.1 radius (the third most distant match). If the cluster is the correct counterpart, the age would be 1 Myr and mass 42260.0 $\Msun$. However, we do not believe this is a probable match and do not include it in our analysis. 

\cite{2006ApJ...645..264T} find that three of their ULXs may be in or near a star cluster. This corresponds to Src09, Src26, Src36 in our nomenclature. They suggest a fourth source, corresponding to Src13 may be close to but unassociated with a star cluster. We do not find a statistically significant match within the 1.1 arcsecond radius between this source and anything in the full LEGUS catalog.  We repeat the cautionary note from earlier that on an individual basis, careful astrometry should be considered. 

Given that our ultimate goal is to obtain the maximum possible number of accreting IMBHs to provide a benchmark to compare with a leading theory of the formation of massive black holes, despite the difficulties of determining optical counterparts in crowded fields, this cross matching serves to act as a naive upper limit for comparison. However, we again caution against the use of any individual counterpart in further studies without performing careful astrometry. 

With these caveats in mind, we found that most of the potential cluster candidate counterparts to bright X-ray sources in M51 had cluster masses less than 1000 $\Msun$, four had masses between 1000 $\Msun$ and 10,000 $\Msun$. Three had masses over 10000 $\Msun$ (Figure \ref{Fig:mass_age2}). Nine are young with ages less than 4 Myr, one has an age estimate of 6 Myr, four are between 40 and 200 Myr, and one is 2000 Myr (Figure \ref{Fig:mass_age}). 

The ages are significant; while it is well known that globular clusters do not have a dense intracluster medium (ICM, \citealt{2001ApJ...557L.105F}), young clusters with ages $<$ 4 Myr may still have an ICM for a black hole to accrete from. A number of studies including \cite{2014MNRAS.443.3594B, 2015MNRAS.449.1106H, 2015MNRAS.448.2224C,2019MNRAS.490.4648H} suggest that young massive clusters lose their gas within 2-4 Myr of their lives, so clusters up to 4 Myr may still have enough gas for a massive black hole to accrete from, whereas the X-rays produced by the clusters older than 4 Myr must be due to accretion from a binary companion to the compact object. The fast clearing time is also consistent with the anti-correlation with 3.3 $\mu$m polycyclic aromatic hydrocarbon non-detection by \cite{2023ApJ...944L..26R}. 

\begin{figure*}
    \centering
    \includegraphics[width=7in]{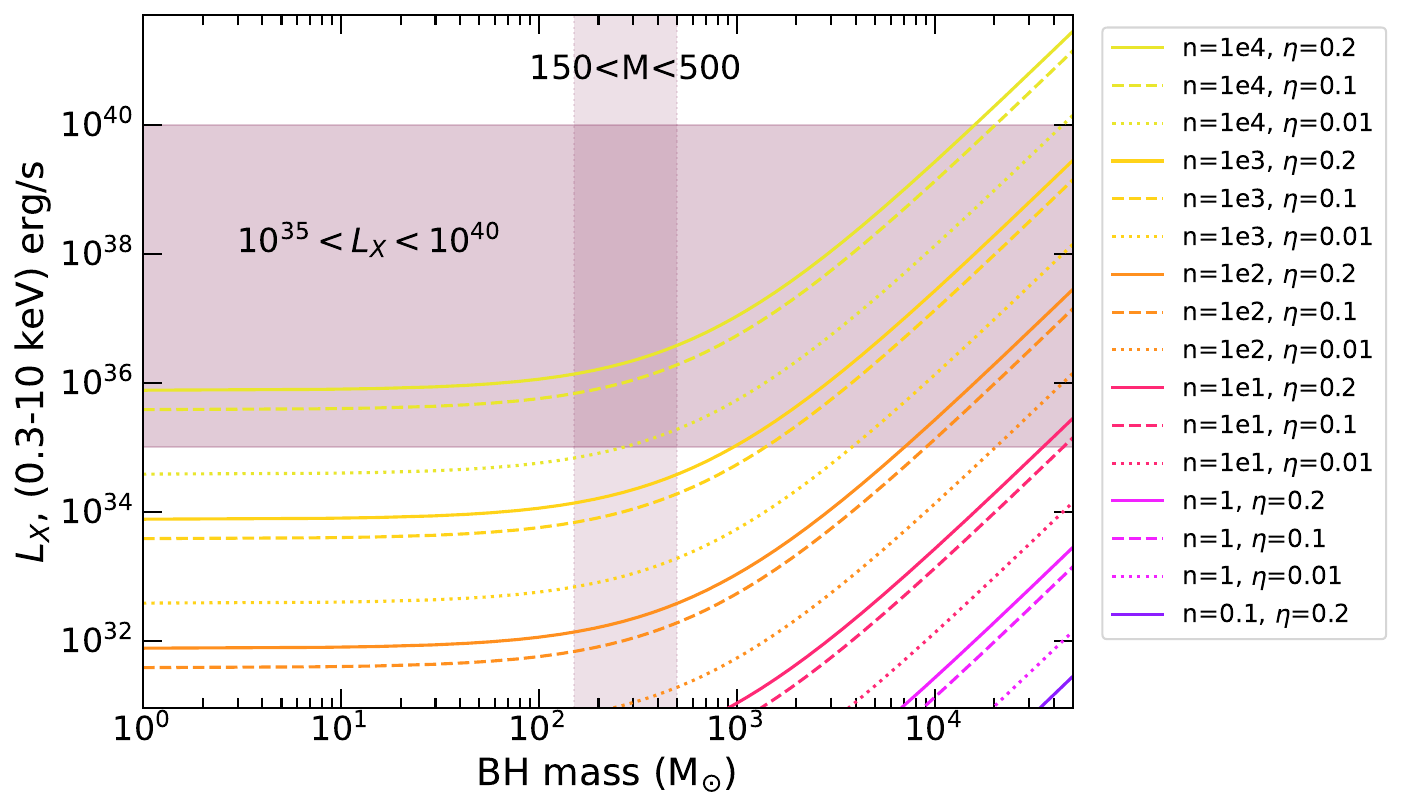}
    \caption{Predicted X-ray luminosity for embedded gas of different densities (from n=0.1 which is more typical of an older globular cluster to n=$10^4$ cm$^{-3}$) and different accretion efficiencies (1\%, 10\% and 20\%) assuming Bondi-Hoyle-Littleton accretion. At the distance of M51, unless the gas is extremely dense ($10^4$ cm$^{-3}$), detecting IMBHs with mass less than $10^3 \Msun$ in the X-ray is difficult. For BHs with mass $>$ $10^{3}\Msun$, densities above 1 can be observable in X-ray, but we would only expect radio emission in the cases with 1\% accretion efficiency, which would require either the higher density ($10^4$ cm$^{-3}$) or a black hole more massive than $5\times 10^4 \Msun$. }
    \label{fig:densitylx}
\end{figure*}
Notably, as shown in Figure \ref{fig:age_mass}, the majority of probable matches to clusters with ages less than 4 Myr are all to cluster candidates below 1000 $\Msun$. The two most massive cluster candidate counterparts with ages less than 4 Myr and masses above 1000 $\Msun$ are src31 and src36. Both have X-ray luminosities above $10^{39}$ erg/s, and low probability matches (25\% and 12\% respectively), and matches with similar probabilities are class 4 contaminants. Therefore, it is quite unlikely that either of these sources are associated with the nearby star cluster.

\subsection{Comparison to star cluster simulations}

While it is non-trivial to compare observations to theory due to the number of difficulties in interpreting the nature of the X-ray sources, given the wealth of theory, we naively make an attempt to reconcile our observations with current predictions by simulations. For example, \cite{2004Natur.428..724P} show how massive black holes can form via runaway collisions in a young cluster. 

In newer simulations by \cite{2021MNRAS.507.5132D}, the most massive black hole formed is just under 500 $\Msun$ in a low metallicity cluster, with the majority of IMBHs having masses between 100$\Msun$ and 200 $\Msun$, although these results are sensitive to the initial conditions of the simulation. They found that under 1\%  star clusters with masses between 1000 and 5000 $\Msun$ form an IMBH,  but 8\% of star clusters with masses between $10^4 \Msun$ and $5\times10^4 \Msun$ produce an IMBH. For the lowest mass clusters, \cite{2021MNRAS.507.3612R} finds that up to 85\% of the IMBHs are ejected from the parent cluster. \cite{2022MNRAS.517.2953T} finds that for clusters with masses between 500 and 800 $\Msun$ less than 0.01\% of the black holes they form are still bound to their low mass parent cluster.

One of the major challenges of comparing X-ray observations of potential black holes to theory is not knowing how many of the IMBHs predicted from the simulations will be accreting from a stellar companion, and thus producing observable X-rays. However, if the ICM is dense enough for the black hole to accrete from, we might expect to see more X-ray sources produced by clusters with ages less than 4 Myr. We therefore separate out clusters with ages of less than 4 Myr and assess whether they host more X-ray sources than the older clusters. 

For star clusters with ages $<$ 4 Myr, there are almost 800 star clusters in the 500-800 $\Msun$ mass range,  just under 1000 with masses in the 1000-5000$\Msun$ mass range (Figure \ref{fig:age_mass}). $\sim$ 130 clusters fell into  $1\times 10^{4}-5\times 10^{4}$$\Msun$ mass range. If we assume that IMBHs in clusters with ages $<$ 4 Myr can efficiently accrete off of the ICM and produce X-rays above some amount, we might expect to see 10 X-ray sources (1\%) in 1000-5000$\Msun$ mass range and a further ten (8\%) in  $1\times 10^{4}-5\times 10^{4}$$\Msun$ mass range. We see no bright X-ray sources with high probability matches and ages less than 4 Myr in these mass ranges, and two low probability matches. Our search for X-ray faint sources in the longest observations (Table \ref{table:appendix}) yielded three high probability matches with $L_X$ between $9\times 10^{36}$ erg/s and $2\times 10^{37}$ erg/s to clusters with ages less than 4 Myr and masses between 500$\Msun$ and 5000 $\Msun$. 

\subsection{Implications of radio observations}

Our limiting 10GHz radio luminosity is $\sim 10^{34}$ erg/s. If we assume a point source with a flat spectrum, then the 5GHz limiting radio luminosity is $7\times 10^{33}$ erg/s. Using the assumptions of the fundamental plane and equation (8) from \cite{Gultekin19}, our current radio sensitivity limits mean that we are only able to detect radio emission from IMBHs with masses $10^4 \Msun$ or above. This is sensible: Figure \ref{fig:accretioneff} shows IMBHs with masses below $10^3 \Msun$ will not have low efficiency accretion for the very brightest X-rays ($10^{38}$ erg/s or above) where we would expect to see radio.

With the advent of next generation radio facilities like the next generation VLA\footnote{\url{https://ngect.nrao.edu/}}, we can push the limiting radio luminosities down to $7\times 10^{32}$ erg/s which corresponds to an increased sensitivity to black holes of masses down to $10^3 \Msun$. Our estimates for detection of IMBHs at 10 Mpc are roughly consistent with estimates of radio emission from IMBHs at 17 Mpc by \cite{2021ApJ...918...18W}. This is also true for the Square Kilometre Array, the most recent version of the SKA-Mid sensitivity calculator \footnote{\url{https://sensitivity-calculator.skao.int/mid}} suggests that a 1 hour exposure will produce a limiting RMS of 550 nJy. Our estimates are consistent with similar estimates for IMBHs at 20 Mpc by \cite{2024arXiv240902893K}.
\subsection{Speculation on detectability of accreting IMBHs in young massive star clusters}
While it is very difficult to interpret the nature of the X-ray source in the absence of radio detections, we can explore the nature of X-ray emission from M51's star clusters, and if any at all are likely to be IMBHs. We first note that X-rays of the luminosities we discuss here can be produced by multitude of objects: IMBHs accreting under the assumptions of the fundamental plane, low-luminosity accretion from an IMBH, or super-Eddington accretion from a stellar mass black hole or neutron star.

We attempt to discriminate between these scenarios, but acknowledge that in the absence of a secure radio detection that can be linked to the X-ray using the fundamental plane, it is impossible to truly classify the nature of the compact object. 
\cite{2023A&A...671A.149F,2024A&A...684A.124F} show that no LMXBs persistently exceed $10^{38}$ erg/s in our Galaxy, and that only three HMXBs exceed $10^{38}$ erg/s. Indeed, the HXMB X-ray luminosity function (XLF) clearly peaks below $10^{36}$ erg/s. We know that the XLF varies depending on Galaxy type, and that ULXs with both high and low mass companion stars can be observed outside the Milky Way \citep{2014ApJ...789...52L,2016ApJ...818...33P}, but this already highlights the unique nature of these X-ray bright sources, whether they are IMBHs or stellar-mass compact objects. While we cannot appeal to our understanding of Milky Way bright X-ray sources to help us interpret those found in M51, we can make some naive speculations to try to better align IMBH predictions from theory and what we are able to observe. 

Given the current observational constraints, what does observing an IMBH look like in star clusters at M51's distance? 
There are roughly 2500 star clusters in M51 with masses between $10^4 \Msun$ and $5 \times 10^4 \Msun$. If up to 8\% of these clusters can host IMBHs, that translates to 200 possible IMBHs. Not all of these are necessarily in a binary accreting at high enough rates to produce observable X-rays at the distance of M51. Currently, we see two high-probability X-ray sources with $L_X > 10^{35}$ erg/s matching to clusters that match this criterion, which means that if both of these are IMBHs, then only 1\% of these IMBHs can be detected in X-ray in present day clusters.  

However, in the case of clusters with ages less than 4 Myr, it may be possible that there is enough intercluster medium and winds from supergiant stars which have not yet been swept out. Figure \ref{fig:densitylx} shows the predicted X-ray luminosities for Bondi-Hoyle-Littleton accretion for a range of black hole masses. We follow the same assumptions as \cite{2024ApJ...961...54P} except that while they fix the number density to 0.2 for globular clusters, we test it for a range of densities from 0.01 to $10^4$ cm$^{-3}$.  This is only a simplistic assessment, and we refer the reader to \cite{2021ApJ...914..109H} for more complicated interactions like irradiation-driven winds.

For black holes in the mass range of 150-500 $\Msun$, we will be able to detect them in X-ray at these distances if they are accreting from ICM with the number density of $10^4$ cm$^{-3}$. For more massive black holes, we can start detecting them in X-ray at the distance of M51 with lower densities, but for the ones that have radiatively inefficient accretion will be towards the tail end of the X-ray luminosities we are sensitive to, and not necessarily detected in radio. For these young regions, 1-$10^3$ cm$^{-3}$ are reasonable densities to expect in \ion{H}{2} regions, based on those observed in the LMC and SMC \citep{2014ApJ...795..121L}. Higher densities are observed in the ultracompact \ion{H}{2} regions, but this phase is short lived \citep[$<$ 1 Myr; ][]{2002ARA&A..40...27C}. There may be other contributions to increase the number density of the ICM via wind loss through massive stars in the cluster \citep{2020MNRAS.493.1306C}, which would lead towards being able to detect BHs less massive than $10^3 \Msun$ in the X-ray, if not in radio.

 Black holes with masses well above $10^4 \Msun$ are unlikely to form in these cluster systems \citep{2012MNRAS.423.1309M}. Even if they were likely to form, detecting them would be challenging. As discussed previously, if the $10^4 \Msun$ IMBH could form at a young enough age to be embedded in an intercluster medium ($<4$ Myr), then accretion from the dense gas could give rise to X-ray and radio emission detectable with current facilities. 

Accretion by a $\sim 10^4\ \Msun$ IMBH is known to generate X-ray and compact radio emission, as in the case of the IMBH candidate HLX-1, a hyper-luminous X-ray source with observed spectral state transitions reminiscent of black hole binaries \citep{Webb}. The known episodic ejections produced during high X-ray state in XRBs was observed in HLX-1, although detection in this ejection phase depicted in the fundamental plane relation is known to be relatively difficult for IMBHs \citep{Yang2023}.

Tidal disruption event (TDE) by an IMBH may be another possibility to explain the X-ray emission we see. 
Depending on the assumptions made in the evolution of TDE modelling, accretion onto the disk can remain at a fallback rate that is still super-Eddington at late times, in some cases up to a few to tens of years from the initial disruption \citep{Tang2024}. In this phase, the bulk of the disk emission occurs in the soft X-rays. Very late-rising radio-luminous TDEs have been observed but only for higher black hole masses, as in the case of \cite{Zhang2024}, where a late-rising radio emission proceeded the decay of optical light curve with a delayed soft X-ray flare. In fact, \cite{Cendes2024} found that $40\%$ of all optical TDEs are detected in radio hundreds to thousands of days after discovery. Though these TDEs are due to black holes a few orders of magnitude beyond black hole masses considered for star clusters, nonetheless, it may be worth noting that while such TDEs are possible perhaps at lower black hole masses, no unambiguous radio emission due to TDE by an IMBH of $\sim 10^4\ \Msun$ have been reported. Furthermore, the predicted disruption rate of TDEs are generally quite low and can vary up to a few factors due to considerable uncertainties in the modelling of TDEs around IMBHs, e.g., \cite{Tang2024,Rizzuto_2023}, and depends sensitively on the stellar distribution around the black hole set by the relaxation timescale. For low density stellar environments, relaxation takes longer, and the continued supply of stars into the lost cone orbits required for disruptions is therefore less frequent and thus TDE rate is much lower. 

In this case, we are unlikely to be sensitive to tidally disrupting IMBHs because these events are extremely rare, and the observing cadence by \textit{Chandra} is unlikely to constrain signatures of TDE in X-ray.

\section{Summary and Conclusions} \label{sec:summary}
We leveraged existing star cluster catalogs from LEGUS and archival X-ray observations from \textit{Chandra} to search for bright X-ray point sources associated with or possibly hosted by star clusters, using a Bayesian cross matching algorithm. We performed model fitting of the HST observations to classify the cluster candidates based on their structural parameters and found that 17 of the brightest X-ray sources likely matched to a cluster candidate, with 11 having greater than 50\% probability of matching. 14 of the lower luminosity sources were high probability matches to a star cluster candidate. 

We compared the X-ray luminosity to the age and mass estimates for the clusters and found that the majority of the X-ray sources were affiliated with young, low mass clusters. We searched for potential radio counterparts to these and did not find any emission above a radio luminosity of  $10^{34}$ erg/s at 10GHz, except for Src03 which had a 70$\mu$ Jy radio counterpart at 10GHz. This radio counterpart was also detected by \cite{2007AJ....133.2559M}, who classify it as a compact H$\alpha$ source. Because the X-ray spectrum of Src03 is soft \citep{2023ChJPh..83..579S}, we thus do not find that the radio emission could be suggestive of an IMBH under the assumptions of the fundamental plane.

We made some initial comparisons to predictions by simulations and found that for star clusters with masses between $10^4\Msun$ and $5\times10^4 \Msun$, only four at most matched to the brightest X-ray sources (i.e. those most likely to be IMBHs), with two being low-probability matches. For the 2500 clusters in this mass range, theory predicts that 8\% or 200 should have IMBHs at some point in their lifetime. This discrepancy implies that only very low fraction of IMBHs are could be producing X-ray emission detectable at the distances of M51. 

Although with the X-ray detections alone, we cannot distinguish X-ray binaries and ULXs with a stellar mass compact object from a bonafide IMBH, we speculate on the observability of IMBHs in both X-ray and radio, at the distances of M51. For clusters with ages $<$ 4 Myr, it is possible that a massive IMBH could produce sufficient X-ray emission to be observed from only the ICM and in the absence of a companion star. While we would not be sensitive to any radio emission from BHs $\lesssim 10^4\Msun$ under these conditions, next generation radio facilities like the Square Kilometre Array and the next generation VLA would be able to detect radio emission from BHs $\gtrsim 10^3\Msun$.

 \begin{acknowledgments}
We thank the referee for helpful comments that greatly improved the manuscript.  The authors thank Jillian Bellovary, Rupali Chandar, Angiraben Mahida and Michela Mapelli for helpful discussion.
 KCD acknowledges support for this work
provided by NASA through the NASA Hubble Fellowship grant
HST-HF2-51528 awarded by the Space Telescope Science Institute, which is operated by the Association of Universities for Research in Astronomy, Inc., for NASA, under contract NAS5–26555.
EWK acknowledges support from the Smithsonian Institution as a Submillimeter Array (SMA) Fellow.

 \end{acknowledgments}
\vspace{5mm}
\facilities{VLA, HST, Chandra}

\software{astropy \citep{Robitaille13}, CASA \citep{CASA2022}, matplotlib \citep{Hunter07}, NumPy \citep{harris20}, pandas \citep{Mckinney10}, NWAY \citep{2018MNRAS.473.4937S}, CIAO \citep{Fruscione06}}

\bibliography{references}{}
\bibliographystyle{aasjournal}

X-ray luminosities and significance from deep observations matched to most probable non-contaminant matches in the LEGUS catalog. We exclude sources away from the galaxy center, as they have a higher probability of being background galaxies.

\begin{sidewaystable}
\centering
\caption{X-ray and Star Cluster properties of low luminosity X-ray sources with high probability matches to star clusters.}
\small
\label{table:appendix}
\begin{tabular}{llrrrrrrrrrrrrrrrrrr}

Chandra ID & HST ID & HST R.A. \&Dec. & F555W & F814W  & C.I. & Age & Mass &  1 &  2 & Sep. & p$_i$ & Sig. & $L_X$ \\
&&&&&&Myr&$\Msun$&&&arcsec&\%&&$10^{36} erg/s$ \\  \hline  \hline
2CXO J133001.4+471157 & 16982 & 13:30:01.37+47:11:57 & 23.70 $\pm$ 0.05 & 22.07 $\pm$ 0.05 & 1.5 &  $<$1&  $<$84 & 0& 3 & 0.71 & 100 & 9.0 & 8.5 $^{+2.5}_{-3.3}$ \\
2CXO J132942.0+471118 & 10925 & 13:29:42.10+47:11:19 & 23.46 $\pm$ 0.05 & 23.08 $\pm$ 0.05 & 1.4 & 400 & 6317    & 0 &4 & 0.42 & 61 & 15.7 & 18.4 $\pm$ 3.4 \\
2CXO J132953.9+470923 & 2135 &  13:29:53.95+47:09:23 & 21.84 $\pm$ 0.03 & 21.83 $\pm$ 0.04 & 1.6 & $<$1&  $<$84  & 0 &4& 0.45 & 100 & 11.7 & 11.3 $^{+1.64}_{-2.04}$ \\
2CXO J132940.9+471139 & 14091 & 13:29:40.84+47:11:39 & 22.91 $\pm$ 0.03 & 22.76 $\pm$ 0.05 & 1.6 & $<$1&  $<$84  & 0 &2& 0.81 & 100 & 12.8& 15.0 $\pm 3.2$  \\
2CXO J132956.0+471350 & 28079 & 13:29:56.07+47:13:50 & 24.28 $\pm$ 0.06 & 23.36 $\pm$ 0.07 & 1.7 & 1 & 1400      & 0 &2& 0.26 & 100 & 13.1 & 17.5 $^{+3.04}_{-3.46}$\\
2CXO J132955.3+471355 & 28421 & 13:29:55.30+47:13:55 & 23.86 $\pm$ 0.06 & 22.92 $\pm$ 0.06 & 1.7 & 200 & 6827    & 0 &0& 0.48& 100 &  16.1 & 22.2 $\pm 3.7$  \\
2CXO J132958.7+471030 & 6559 &  13:29:58.72+47:10:30 & 23.60 $\pm$ 0.05 & 21.05 $\pm$ 0.04 & 1.6 & $<$1&  $<$84  & 0 &3& 0.20 & 69 & 19.4 & 18.2 $\pm 2.9$ \\
2CXO J132950.3+471322 & 25585 & 13:29:50.36+47:13:22 & 23.91 $\pm$ 0.06 & 22.74 $\pm$ 0.05 & 1.5 & 40 & 4301     & 0 &4& 0.46 & 100 & 18.8 & 28.1 $\pm$ 3.9 \\
2CXO J132944.0+471156 & 16848 & 13:29:44.13+47:11:56 & 24.03 $\pm$ 0.08 & 23.67 $\pm$ 0.13 & 1.5 & 100 & 1878    & 0 &0& 0.56 & 50 & 21.1 & 27.2 $\pm$ 3.8 \\
2CXO J132942.5+471042 & 7928 &  13:29:42.55 +47:10:43 & 23.56 $\pm$ 0.06 & 23.28 $\pm$ 0.12 & 1.4 & 200 & 4260   & 0 &0& 0.83& 100 & 24.0 & 29.8 $\pm$ 4.1 \\
2CXO J132952.7+471244 & 21961 & 13:29:52.72+47:12:45 & 23.99 $\pm$ 0.06 & 23.38 $\pm$ 0.09 & 1.5 & 10 & 316.7    & 0 &4& 0.68& 100 & 30.6 & 40.8 $\pm$ 4.7 \\
2CXO J132949.6+470910 & 1490 &  13:29:49.60+47:09:10 & 23.95 $\pm$ 0.06 & 22.97 $\pm$ 0.05 & 1.5 & 40 & 2663     & 0 &2& 0.14 & 70 & 31.2 & 35.7 $\pm$ 4.4 \\
2CXO J132934.9+470934 & 2633 &  13:29:34.92+47:09:34 & 23.46 $\pm$ 0.06 & 22.65 $\pm$ 0.07 & 1.6 & 100 & 5092    & 0 &0& 0.51 & 100 & 29.2 & 52.8 $\pm$ 6.6 \\
2CXO J133000.7+471212 & 18754 & 13:30:00.79+47:12:13 & 24.13 $\pm$ 0.06 & 23.43 $\pm$ 0.09 & 1.5 & $<$1&  $<$84  & 0 &0& 0.89 & 100 & 32.9& 38.2 $\pm$ 4.3 \\
2CXO J133004.0+471003 & 4772 &  13:30:04.08+47:10:03 & 23.30 $\pm$ 0.04 & 22.77 $\pm$ 0.04 & 1.7 & $<$1&  $<$84  & 0 &0& 0.21& 67& 37.2 & 40.2 $\pm$ 4.1 \\
2CXO J133004.5+470949 & 3641 &  13:30:04.47+47:09:49 & 21.73 $\pm$ 0.03 & 21.21 $\pm$ 0.03 & 1.6 & $<$1&  $<$84  & 0 &0& 0.36 & 100 & 37.1 & 34.1 $\pm$ 3.9 \\
2CXO J133010.0+471328 & 25879 & 13:30:10.04+47:13:27 & 21.96 $\pm$ 0.03 & 19.89 $\pm$ 0.03 & 1.8 & 3000 & 1081000& 1 &1& 0.28 & 100 & 9.0 & 8.7 $\pm$ 2.5 \\
2CXO J132937.9+470832 & 333 &   13:29:37.98+47:08:32 & 20.68 $\pm$ 0.02 & 19.40 $\pm$ 0.03 & 1.6 & 200 & 324500  & 1 &1& 0.31 & 55 & 29.8 & 108 $\pm$ 14.0 \\
2CXO J132954.9+471102 & 9341 &  13:29:55.02+47:11:03& 21.05 $\pm$ 0.03 & 20.95 $\pm$ 0.04 & 1.4 & 4 & 5456       & 1 &1& 0.83 & 100 & 34.4 & 39.8 $\pm$ 4.2\\
2CXO J132942.2+471046 & 8237 &  13:29:42.29+47:10:46 & 23.04 $\pm$ 0.04 & 22.37 $\pm$ 0.05 & 1.7 & 200 & 9632    & 2 &2& 0.66& 100 & 23.9 & 29.8 $\pm$ 4.1 \\  \hline
\end{tabular}
\end{sidewaystable}
\end{document}